\documentclass{article}

\usepackage[preprint]{neurips_2026}

\usepackage[utf8]{inputenc} 
\usepackage[T1]{fontenc}    
\usepackage{hyperref}       
\usepackage{url}            
\usepackage{booktabs}       
\usepackage{amsfonts}       
\usepackage{nicefrac}       
\usepackage{microtype}      
\usepackage{xcolor}         

\usepackage{amsmath}
\usepackage{amsthm}
\usepackage{graphicx}
\usepackage{datetime}
\usepackage{multicol}
\usepackage{multirow}

\title{Dynamic Hypergraph Representation Learning for Multivariate Time Series without Prior Knowledge}

\author{%
  Marco Gregnanin \\
  IMT School for Advanced Studies Lucca\\
  KU Leuven \\
  \texttt{marco.gregnanin@imtlucca.it} \\
   \And
   Johannes De Smedt \\
   KU Leuven \\
   \AND
   Giorgio Gnecco \\
 IMT School for Advanced Studies Lucca\\
   \And
   Maurizio Parton \\
  University of Chieti-Pescara \\
}

\begin{document}

\maketitle

\begin{abstract}
Hypergraphs have the capacity to capture higher-dimensional relationships among entities across various domains, making them a subject of growing interest within the research community for understanding the structure and dynamics of complex systems. However, a key challenge is the derivation of hypergraph representations from time series data in situations where the structure of the hypergraph is limited or absent.
In this study, we propose a model that constructs a dynamic hypergraph representation for multivariate time series without relying on prior knowledge of the data. This is achieved by applying community detection to the time series and transforming the resulting communities, obtained through an attention mechanism, into a hypergraph using a clique-based technique. Hypergraph representations are derived from different time series datasets, and the resulting hypergraphs are then used by a Dynamic Hypergraph Attention Convolution Network (DHACN) for multivariate time series predictions. This research advances the field of hypergraph representation by introducing a novel approach that is better suited to uncover high-order relationships without prior knowledge. 
\end{abstract}

\section{Introduction}
Hypergraphs are a generalization of graphs, where a node can be connected to multiple nodes through hyperedges, whereas in traditional graphs, nodes can only be connected pairwise \citep{bretto2013hypergraph}. This ability to connect multiple nodes allows hypergraphs to capture higher-order relationships, which has attracted significant attention from researchers. These higher-order relationships often contain crucial information and characteristics regarding the behavior of nodes within a network. A critical challenge, however, lies in how to derive a hypergraph representation that effectively captures these higher-order relationships within data \citep{antelmi2023survey}. 
Typically, hypergraph representation relies on pre-existing knowledge about the dataset in question, as seen in applications such as citation networks \citep{feng2019hypergraph}, recommendation systems \citep{li2013news}, and traffic forecasting \citep{yi2020hypergraph}. Alternatively, it can be derived from clear data attributes that can be generalized, such as in social networks \citep{gao2022hgnn}.
Hypergraph representations in time series contexts are often derived based on prior knowledge, such as stock market sector classifications \citep{Sawhney2021sthan} or traffic forecasting data \citep{nan2024collaborativehypergraph}. However, deriving a hypergraph representation from multivariate time series without domain knowledge poses significant challenges. This is because it requires clustering the time series data \citep{wang2024dynamic}, which is non-trivial due to the temporal dependencies within the data. Traditional clustering techniques, such as K-Nearest Neighbors (KNN) \citep{peterson2009k}, may neglect these temporal dependencies due to the distance metrics typically used for clustering. \\
In this research, we investigate how to derive dynamic hypergraph representations from multivariate time series without relying on prior knowledge of the data. We propose two methods for deriving these representations based solely on the time series data itself and on Complex Network Theory \citep{barabasi2016network}, and utilize a modularity optimization approach \citep{fortunato2010community} to detect clusters within the time series. 
In the first approach, we filter the time series correlation matrix using the Random Matrix Theory (RMT) approach \citep{mantegna1999introduction}, and then use the filtered matrix to identify clusters among the time series. RMT examines the statistical properties of large random matrices and provides insights into their spectral characteristics. 
The second approach aims to capture inter-series correlations directly from the time series data. Here, we propose to derive the graph representation of the time series using a self-attention mechanism \citep{cao2020spectral}, followed by 
detecting clusters from the self-attention graph presentation.  
Our objective is to derive a hypergraph representation that preserves the underlying network structure of the multivariate time series, and generates task-specific hypergraphs in a data-driven manner.
Finally, we explore these higher-order relationships by applying a Hypergraph Attention Neural Network (HANs) \citep{bai2021hypergraph}, combined with a Long Short-Term Memory (LSTM) network \citep{hochreiter1997long} enhanced by an attention mechanism \citep{vaswani2017attention}, to solve prediction task across different time series datasets comparing the performance results against various baseline models. \\
Our contributions are as follows: 
\begin{itemize}
    \item The introduction of a novel method for deriving dynamic hypergraph representations from time series data without the need for prior knowledge. This method leverages community detection frameworks applied to a correlation matrix filtered through RMT and to a correlation matrix obtained via a data-driven approach.
    \item The introduction of a novel ensemble model that integrates a Dynamic Hypergraph Neural Network with an LSTM enhanced by an attention mechanism.
    \item A thorough evaluation of our approach on various time series datasets, demonstrating improved performance in regression task compared to baseline models. 
\end{itemize}
The rest of the paper is organized as follow: Section \ref{section:related work} reviews hypergraph theory, community detection in time series, and hypergraph representation learning; Section \ref{section:proposed model} outlines the methodology of the proposed model; Section \ref{section:experimental evaluation} details the experimental setup and presents the results of our proposed method compared to baseline models; and Section \ref{section:conclusion} concludes the paper. 

\section{Related Works} \label{section:related work}

In this section, we present a brief review of the hypergraph and dynamic hypergraph representation learning problem as applied to multivariate time series, along with an overview of modularity optimization techniques applied to time series data.

\noindent \textbf{Hypergraph Representation Learning.} Existing hypergraph learning models aim to generate hypergraphs capable of representing higher-order relations within data. In the field of computer vision, \cite{gao20123d} derived hypergraphs by clustering views of a $3$-D object. In \cite{huang2015learning}, the hypergraph was constructed based on shared attributes between different objects. In \cite{jin2019robust}, hyperedges were derived by solving a ridge regression to remove noise from images. Similarly, \cite{feng2019hypergraph} generated hyperedges using a clique technique, connecting K-nearest neighbor nodes to a selected node in an object classification task.
The Uni-GNN model \citep{huang2021unignn} introduced a generalized framework for transitioning from Graph Neural Networks (GNNs) \citep{scarselli2008graph} to Hypergraph Neural Networks (HGNNs) \citep{bai2021hypergraph}, where the hypergraph was constructed from data attributes. In \cite{gao2022hgnn}, hypergraphs were derived using a combination of methods depending on whether the data had an underlying graph structure. In the first approach, a clique-based technique was used, while in the second, hyperedges were generated based on node attributes and the KNN algorithm. 

\noindent \textbf{Hypergraph Representation Learning for Time Series.} Hypergraph representations for time series data are typically constructed based on prior knowledge of the data. For instance, in \cite{yi2020hypergraph}, the hypergraph representation was derived from network sensor data and subsequently analyzed using a combination of Hypergraph Convolutional Networks (HCNs) and Recurrent Neural Networks (RNNs).
Another example of leveraging prior knowledge can be found in traffic forecasting. In \cite{luo2022directed}, the hypergraph is constructed from a traffic network, and traffic forecasting was performed using directed Hypergraph Attention Networks (HANs). Similarly, in \cite{wang2022multitask}, the hypergraph representation was derived from both bus and metro transportation networks to predict metro flow using HCNs followed by a temporal model. In the context of financial data, \cite{Sawhney2021sthan} derived the hypergraph representation among a basket of stocks based on sector classification and corporate relationships to address a stock returns ranking problem using HANs. However, these approaches construct a fixed hypergraph representation that remains unchanged throughout the entire analysis period.

\noindent \textbf{Dynamic Hypergraph Representation Learning.} To capture the evolving structure of hypergraphs over time, a dynamic hypergraph representation is required, allowing both nodes and hyperedges to change over time. A dynamic hypergraph learning approach was proposed by \cite{zhang2018dynamic}, which focused on learning both label projections and data attributes in the context of $3$-D shape and gesture recognition. In \cite{jiang2019dynamic}, a dynamic hypergraph representation for citation networks and social data was derived by combining k-means clustering and KNN to address a classification problem using Hypergraph Convolutional Networks (HCNs). Similarly, in \cite{kang2022dynamic}, dynamic hypergraph representations were generated and analyzed for various datasets, such as citation networks and traffic flow, using the same approach.

\noindent \textbf{Dynamic Hypergraph Representation Learning for Time Series.} Dynamic hypergraph representations for multivariate time series are typically constructed based on prior knowledge of the data. In \cite{yin2022dynamic,nan2024collaborativehypergraph}, dynamic hypergraph representations were derived from the Stocktwits and Tiingo datasets 
to solve a stock price prediction problem, and from the NYC taxi dataset to predict taxi demand using HGNNs. In \cite{Xia2024sthpan}, dynamic hypergraph representations of stock returns were generated based on company sectors, corporate relationships, and Dynamic Time Warping  
computations to address a stock ranking problem using HANs. Similarly, in \cite{wang2024dynamic}, dynamic hypergraph representations across six different datasets, including wind, temperature, and metro data, were constructed using a KNN approach, combining the graph structure derived from attribute knowledge.
Although these models show improved performance, they do not account for how time series may cluster based on their historical observations and inherent similarities.

\noindent \textbf{Random Matrix Theory and Community Detection.} RMT \citep{mantegna1999introduction,edelman2005random} is a tool frequently employed in econophysics and can be combined with community detection algorithms \citep{barabasi2016network,fortunato2010community} to identify how time series data cluster based on historical observations. In \cite{laloux2000random}, RMT was utilized to study the eigenvalue distribution of a correlation matrix derived from stock returns, while \cite{ye2017distinguishing} applied RMT to remove noise from chaotic time series. Additionally, in \cite{ferreira2016time}, community detection was performed on various time series datasets using different clustering methods, demonstrating the benefits of community detection.
In \cite{MacMahon2015} and \cite{gregnanin2023signaturebased}, community detection was applied to the Standard \& Poor's $500$ (S\&P $500$) stock index by filtering the stock returns correlation matrix with RMT. Their results indicated that the number of identified communities was lower than the sector classification provided by the S\&P index, suggesting that the conventional stock classification does not accurately capture the temporal evolution of the time series.

\section{Proposed Framework} \label{section:proposed model}

\subsubsection{Problem Formulation}
\begin{figure*}[t]
\centering
\includegraphics[width=1\textwidth]{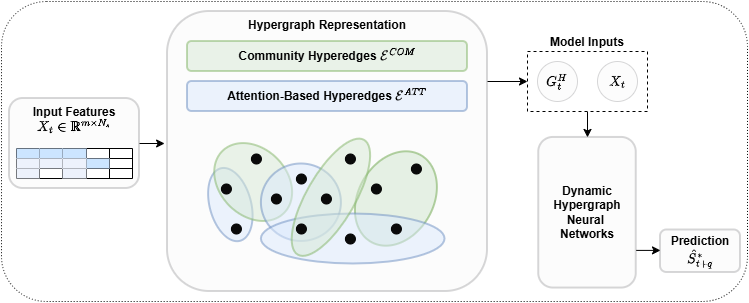}
\caption[Proposed Hypergraph framework.]{An overall pipeline of the proposed approach, showing all the steps needed to arrive at the final outcome starting from $\mathcal{S}$.}
\label{fig:Proposed hypergraph approaches}
\end{figure*}

Let $G^{H} = (\boldsymbol{\mathcal{V}}^{H}, \boldsymbol{\mathcal{E}}^{H}, \mathbf{W}^{H})$ denote a hypergraph, where $\boldsymbol{\mathcal{V}}^{H} = \{v_1, v_2, \dots,$ $v_{N_V}\}$ is the set of nodes and $\boldsymbol{\mathcal{E}}^{H} = \{e_1, e_2, \dots, e_{N_e}\}$ is the set of hyperedges. Each hyperedge $e_i \in \boldsymbol{\mathcal{E}}^{H}$ is defined as a subset of the node set, i.e., $e_i \subseteq \boldsymbol{\mathcal{V}}^{H}$. The diagonal matrix $\mathbf{W}^{H} \in \mathbb{R}^{N_e \times N_e}$ denotes the hyperedge weight matrix.
In this work, we adopt a snapshot-based formulation to model dynamic hypergraphs \citep{antelmi2023survey}. Under this paradigm, a dynamic hypergraph is represented as a sequence of static hypergraph snapshots observed at discrete time instants. Let $\mathcal{T} = \{t_1, t_2, \dots, t_{T_G}\}$ denote the set of timestamps, where $T_G = |\mathcal{T}|$. The dynamic hypergraph is then defined as
\begin{equation*}
    \mathcal{G}^{H} = \{G_{t_{i}}^{H}\}_{t_{i}=1}^{T_{G}}
\end{equation*}
where each snapshot $G_{t_i}^{H} = (\boldsymbol{\mathcal{V}}^{H}_{t_i}, \boldsymbol{\mathcal{E}}^{H}_{t_i}, \mathbf{W}^{H}_{t_i})$ represents the hypergraph at time $t_i \in \mathcal{T}$. Notably, the node set, hyperedge set, and hyperedge weights are allowed to vary across snapshots.
Let $\mathcal{S} = \{S_1, S_2, \dots, S_{N_s}\}$ denote a collection of $N_s$ time series, where each $\mathbf{s}_i$ consists of $T$ temporal observations. For simplicity, we assume that the number of time series, $N_s$, coincides with the number of nodes, $N_V$, in each hypergraph snapshot. Moreover, let $\mathbf{\mathcal{S}}_t = \{S_{1}(t), S_{2}(t), \dots, S_{N_{s}}(t)\}$ denote the set of time series values observed at time $t$. We further assume that the sampling frequency of the time series and the timestamps in $\mathcal{T}$ coincide.\\
The objective of this study is to learn a function $h(\cdot)$ capable of predicting the future $q$ values of a target time series $S^{*} \in \mathcal{S}$ by jointly exploiting the hypergraph representation and historical observations. Specifically, the model takes as input the hypergraph snapshot $G_{t_i}^{H}$ and the past $m$ observations of each time series. Formally, the prediction task is defined as
\begin{equation}
    \hat{S}^{*}_{t+q} = h(\mathbf{X}_t, G_{t_{i}}^{H}) \qquad \forall t\in T, \quad \forall t_{i} \in \mathbf{\mathcal{T}}\,, \label{eq:h_function_hypergraph}
\end{equation}
where $\mathbf{X}_t \in \mathbb{R}^{m \times N_s}$ is the feature matrix collecting the past $m$ observations of the $N_s$ time series in tabular form, namely
\begin{equation*}
    \mathcal{S}_{t-m+1:t} = \left\{S_{1}(t-m+1:t), S_{2}(t-m+1:t), \dots, S_{N_{s}}(t-m+1:t)\right\}\,.
\end{equation*}
The hypergraph snapshot $G_{t_i}^{H}$ is constructed from these same historical observations, i.e., from $\mathcal{S}_{t-m+1:t}$. Consequently, the total number of hypergraph snapshots depends on the length of the time series $T$ and the window size $m$, yielding $T_G = T - m + 1$.\\
An overview of the proposed framework is illustrated in Figure \ref{fig:Proposed hypergraph approaches}. The procedure consists of three main steps: extracting the past $m$ observations from the multivariate time series; constructing the corresponding hypergraph representation $G_{t_i}^{H}$; and jointly leveraging $G_{t_i}^{H}$ and the feature matrix $\mathbf{X}_t$ to predict the future values $\hat{S}^{*}_{t+q}$.

\subsubsection{Dynamic Hypergraph Representation}

In our construction, hyperedges are obtained via community detection over two time-varying dependency estimates: a correlation matrix filtered through Random Matrix Theory (RMT), and a latent dependency matrix learned through a self-attention mechanism. Both mechanisms aim to identify groups of time series that exhibit stronger within-group dependence than cross-group dependence.

\paragraph{Community Hyperedges.} 
Time series assigned to the same community are connected by a hyperedge $e_{t_i,j} \in \boldsymbol{\mathcal{E}}^{\mathrm{COM}}_{t_i}$, where
\begin{equation*}
    \boldsymbol{\mathcal{E}}^{\text{COM}}_{t_{i}}=\{e_{t_{i},1}, e_{t_{i},2}, \dots e_{t_{i},N_{e_{k}}}\}\,,
\end{equation*}
and $N_{e_k}$ denotes the number of detected communities. This approach is motivated by the observation that communities in financial correlation networks often correspond to groups of series that are positively correlated within the group and comparatively negatively correlated across groups \citep{MacMahon2015}.
At each time $t$, we compute an empirical correlation matrix $C_t \in \mathbb{R}^{N_V \times N_V}$ using the past $m$ observations $\mathcal{S}_{t-m+1:t}$. Before performing community detection, we apply RMT-based filtering to remove noise and retain only statistically significant components\footnote{A detailed description of the procedure for performing community detection using Random Matrix Theory is provided in Appendix \ref{Appx:rmt}.}.

\paragraph{Attention-Based Hyperedges.} 
The second mechanism constructs hyperedges from a latent dependency matrix inferred via self-attention. Starting from the input matrix $\mathbf{X}_t \in \mathbb{R}^{m \times N_s}$, we first apply a Gated Recurrent Unit (GRU) layer \citep{cho2014properties} to capture temporal dependencies. Following \cite{zhang2023dive}, the GRU update is:
\begin{align}
    \mathbf{R}_{t} &= \sigma \left(\mathbf{X}_{t}\mathbf{W}_{xr} + \mathbf{Z}_{t-1}\mathbf{W}_{zr} + \mathbf{b}_{r}\right) \,,\nonumber \\
    \mathbf{U}_{t} &= \sigma \left(\mathbf{X}_{t}\mathbf{W}_{xu} + \mathbf{Z}_{t-1}\mathbf{W}_{zu} + \mathbf{b}_{u}\right) \,,\nonumber \\
    \Tilde{\mathbf{Z}}_{t} &= \text{tanh} \left(\mathbf{X}_{t}\mathbf{W}_{xz} + (\mathbf{R}_{t} \odot \mathbf{Z}_{t-1})\mathbf{W}_{zz} + \mathbf{b}_{z}\right) \,,\nonumber \\ 
    \mathbf{Z}_{t} &= \mathbf{U}_{t} \odot \mathbf{Z}_{t-1} + (1 - \mathbf{U}_{t}) \odot \Tilde{\mathbf{Z}}_{t} \,.   \label{eq:gru}
\end{align}
Here $\mathbf{R}_t,\mathbf{U}_t\in\mathbb{R}^{m\times z_h}$ are the reset and update gates, $\widetilde{\mathbf{Z}}_t\in\mathbb{R}^{m\times z_h}$ is the candidate hidden state, and $\mathbf{Z}_t\in\mathbb{R}^{m\times z_h}$ is the hidden state. The matrices $\mathbf{W}_{xr},\mathbf{W}_{xu},\mathbf{W}_{xz}\in\mathbb{R}^{N_V\times z_h}$ and $\mathbf{W}_{zr},\mathbf{W}_{zu},\mathbf{W}_{zz}\in\mathbb{R}^{z_h\times z_h}$ are learnable parameters, $\mathbf{b}_r,\mathbf{b}_u,\mathbf{b}_z\in\mathbb{R}^{1\times z_h}$ are biases, $\odot$ denotes the Hadamard product, and $z_h$ is the dimension of the hidden units.\\
We then define query and key matrices as $\mathbf{Q}_t = \mathbf{Z}_t\mathbf{W}_Q$ and $\mathbf{K}_t = \mathbf{Z}_t\mathbf{W}_K$, where $\mathbf{W}_Q,\mathbf{W}_K\in\mathbb{R}^{z_h\times N_s}$ are learnable. The resulting latent dependency matrix $\mathbf{A}_t\in\mathbb{R}^{N_s\times N_s}$ is computed by
\begin{equation}
    \mathbf{A}_t = \text{softmax} \left( \frac{\mathbf{Q}_{t}^{T}\mathbf{K}_{t}}{\sqrt{N_{V}}} \right) \,. \label{eq:latent_correlation}
\end{equation}
We interpret $\mathbf{A}_t$ as an attention-derived analogue of a correlation matrix and apply the same modularity-based community detection as in \eqref{eq:modularity}\footnote{Defined in Appendix \ref{appx_modularity}.}, replacing $\mathbf{C}^{*}_t$ with $\mathbf{A}_t$. The resulting communities produce 
attention-based hyperedges $e_{t_i,j}\in\boldsymbol{\mathcal{E}}^{\mathrm{ATT}}_{t_i}$, where
\begin{equation*}
    \boldsymbol{\mathcal{E}}^{\text{ATT}}_{t_{i}}=\{e_{t_{i},1}, e_{t_{i},2}, \dots e_{t_{i},N_{e_{k'}}}\}
\end{equation*}
and $N_{e_{k^{'}}}$ is the number of communities obtained from $\mathbf{A}_t$.

\paragraph{Hypergraph Assembly and Weighting}
We form the final hyperedge set by union:
\begin{equation*}
    \boldsymbol{\mathcal{E}}_{t_{i}}^{H} = \boldsymbol{\mathcal{E}}^{\text{COM}}_{t_{i}} \cup \boldsymbol{\mathcal{E}}_{t_{i}}^{ATT} \,.
\end{equation*}
The hypergraph can be represented via an incidence matrix $\mathbf{H}^{H}_{t_i}\in\mathbb{R}^{|\boldsymbol{\mathcal{V}}^{H}|\times |\boldsymbol{\mathcal{E}}_{t_i}^{H}|}$ with entries
\begin{equation*}
    h_{t_{i}}(v, e_{t_{i},j}) = \begin{cases}
        1 & \text{ if } v \in e_{t_{i},j}\,, \\
        0 & \text{ otherwise}\,.
    \end{cases} 
\end{equation*}
Each hyperedge is assigned a positive weight $w_{t_i}(e_{t_i,j})$, stored in the diagonal matrix $\mathbf{W}^{H}_{t_i}\in\mathbb{R}^{|\boldsymbol{\mathcal{E}}^{H}_{t_i}|\times |\boldsymbol{\mathcal{E}}^{H}_{t_i}|}$. In particular, we define
\begin{equation*}
    w_{t_{i}}(e_{t_{i},j}) = \sum_{v_h,v_k \in e_{t_{i},j}, h \neq k} | a_t(v_h, v_k)| \,,
\end{equation*}
where $a_t(v_h,v_k)$ is the $(h,k)$-th entry of $\mathbf{A}_t$ in \eqref{eq:latent_correlation}. Finally, the vertex degree $d_{t_i}(v)$ and the hyperedge degree $\delta_{t_i}(e_{t_i,j})$ are given by
\begin{align*}
    d_{t_{i}}(v) &= \sum_{e_{t_{i},j}\in \boldsymbol{\mathcal{E}}^{H}_{t_{i}}}  w_{t_{i}}(e_{t_{i},j})h_{t_{i}}(v, e_{t_{i},j})\,, \\
    \delta_{t_{i}}(e_{t_{i},j}) & = \sum_{v \in \mathbf{\boldsymbol{V}}^{H}_{t_{i}}}h_{t_{i}}(v, e_{t_{i},j})\,.
\end{align*} 
Let $\mathbf{D}^{v}_{t_i}\in\mathbb{R}^{|\boldsymbol{\mathcal{V}}^{H}_{t_i}|\times |\boldsymbol{\mathcal{V}}^{H}_{t_i}|}$ and $\mathbf{D}^{e}_{t_i}\in\mathbb{R}^{|\boldsymbol{\mathcal{E}}^{H}_{t_i}|\times |\boldsymbol{\mathcal{E}}^{H}_{t_i}|}$ denote the corresponding diagonal matrices of vertex degrees and hyperedge degrees at time $t_i\in\mathcal{T}$, respectively.

\subsubsection{Dynamic Hypergraph Neural Networks}
\begin{figure*}[t]
\centering
\includegraphics[width=1\textwidth]{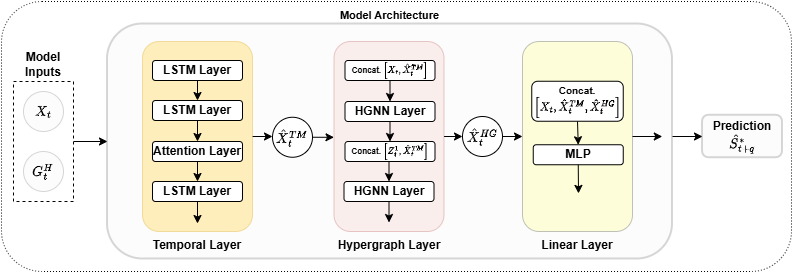}
\caption[Proposed Hypergraph Neural Networks Architecture.]{Proposed Hypergraph Neural Networks Architecture.}
\label{fig:Hypergraph architecture}
\end{figure*}
Figure \ref{fig:Hypergraph architecture} shows the model architecture of the proposed Dynamic Hypergraph Neural Networks (DHNN). The model takes ad input the feature matrix $X_t \in \mathbb{R}^{m \times N_s}$ and the hypergraph representations $G_t^{H}$, in order to predict the future $q$ values of the target time series $S^{*}_{t+q}$, as defined in equation \ref{eq:h_function_hypergraph}. 
As shown in Figure \ref{fig:Hypergraph architecture}, the proposed hypergraph neural network architecture is composed of three different layers; a \emph{Temporal Layer}, which captures the temporal dependencies; a \emph{Hypergraph Layer}, which models the high-order relationship; and a \emph{Linear Layer}, which integrates the learned representation to produce the final forecast. 

\paragraph{Temporal Layer}
The temporal layer takes as input $X_t \in \mathbb{R}^{m \times N_s}$, and produces a temporal representation $\widetilde{X}^{TM}_t \in \mathbb{R}^{m \times N_s}$. This module combines a stacked LSTM layer with a multi-head self-attention mechanism, followed by a fully connected transformation. Its main objective is to extract a compact latent representation of the input sequence while preserving both short-term and long-term temporal structures. \\
First, the input is processed through two stacked LSTM layers:
\begin{equation*}
    H^{(1)}_t = LSTM_1(X_t)\,, \ \ \ \ \ \ H^{(2)}_t = LSTM_2(H^{(1)}_t)\,,
\end{equation*}
where $H^{(1)}_t \in \mathbb{R}^{m \times h_1}$, $H^{(2)}_t \in \mathbb{R}^{m \times d}$, with $h_1$ the hidden dimension of the first LSTM layer, $d=h_2\cdot n_{head}$ is the hidden dimension of the second LSTM layer with $h_2$ representing the number of neurons and $n_{head}$ the number of attention heads. 
A multi-head self-attention mechanism is then applied: 
\begin{equation*}
    \widetilde{H}_t = \text{MultiHeadAttention}(H^{(2)}_t)
\end{equation*}
where $\widetilde{H}_t \in \mathbb{R}^{m \times d}$ which allows the model to assign different importance weights to different time steps. We can define the attention score using the scaled dot-product as: 
\begin{equation*}
    A^{(k)}_t = \frac{Q^{(k)}_t(K^{(k)}_t)^{T}}{\sqrt{d_k}}\,, \ \ \ \ \ \ \ 
    \forall k = 1, \dots, n_{head}
\end{equation*}
where $Q^{(k)}_t = H^{(2)}_tW_{Q}^{(k)}$ are the queries, $K^{(k)}_t = H^{(2)}_tW_{K}^{(k)}$  the keys, and $W_{Q}^{(k)},W_{K}^{(k)} \in \mathbb{R}^{d \times d_k}$ are learnable projection matrices with $d_k = d/n_{head}$. 
A softmax function is applied row-wise to obtain attention weights:
\begin{equation*}
    \alpha^{(k)}_t = \text{softmax}\left(A^{(k)}_t\right)
\end{equation*}
Then, the output of the $k$-th head is:
\begin{equation*}
    \text{head}^{(k)}_t = \alpha^{(k)}_t V^{(k)}_t
\end{equation*}
where $V^{(k)}_t = H^{(2)}_tW_{V}^{(k)}$ are the values with $W_{V}^{(k)} \in \mathbb{R}^{d \times d_k}$ learnable matrix weights. Finally, the outputs of the heads are concatenated as:
\begin{equation*}
    \text{MultiHead}\left( H^{(2)}_t\right) = \|_k^{n_{head}}\text{head}^{(k)}_t \in \mathbb{R}^{n \times d}
\end{equation*}
A final linear transformation is applied: 
\begin{equation*}
    \widetilde{H}_t = \text{MultiHead}\left( H^{(2)}_t\right) W_0
\end{equation*}
where $W_0 \in \mathbb{R}^{d \times d}$ is a learnable projection matrix.\\
The resulting sequence is further processed by a third LSTM layer:
\begin{equation*}
    H^{(3)}_t = LSTM_1\left(\widetilde{H}_t\right)\,,
\end{equation*}
Before obtaining the temporal representation $\widetilde{X}^{TM}_t$, we apply a fully connected layer at $H^{(3)}_t$. 

\paragraph{Hypergraph Layer}
The hypergraph layer takes as input $X_t \in \mathbb{R}^{m \times N_s}$, $\widetilde{X}^{TM}_t \in \mathbb{R}^{m \times N_s}$, and the hypergraph representation $G^{H}_t$, and produces the output $\widetilde{X}^{HG}_t \in \mathbb{R}^{m \times N_s}$. 
The purpose of this module is to capture higher-order structural dependencies among variables through a learnable hypergraph representation. \\
First, the temporal representation $\widetilde{X}^{TM}_t$ is concatenated with the initial input $X_t$. Then, it is processed through a sequence of convolutional hypergraph layers. The first hypergraph convolution is defined as:
\begin{align*}
    Z^{(0)}_t &= \left[X_t \| \widetilde{X}^{TM}_t\right]\\
    Z^{(1)}_t &= \sigma \left( \text{HGNN}_{1}\left( Z^{(0)}_t, G^{H}_t \right) \right) 
\end{align*}
where $\sigma (\cdot)$ is an Exponential Linear Unit (ELU)\footnote{The ELU activation function is defined as:
\begin{equation*}
    ELU(x) = \begin{cases}
        x & \text{ if } x>0 \\
        \alpha (\exp{(x)} - 1) & \text{ if } x\le 0
    \end{cases}
\end{equation*}
with $\alpha > 0 $} activation function \citep{clevert2016elu}.
Then, the output is concatenated with the temporal representation:
\begin{equation*}
    Z^{(\text{cat})}_t = \left[Z^{(1)}_t \| \widetilde{X}^{TM}_t\right]
\end{equation*}
Then a linear transformation is applied:
\begin{equation*}
    \hat{Z}^{(1)}_t = W_1 Z^{(\text{cat})}_t 
\end{equation*}
where $W_1$ is a learnable weight matrix. Finally, a second hypergraph convolutional layer refines the representation:
\begin{equation*}
    \widetilde{X}^{\text{HG}}_{t} = \sigma \left( \text{HGNN}_{2}\left( \hat{Z}^{(1)}_t, G^{H}_t \right) \right) 
\end{equation*}
Note that a generic hypergraph convolutional layer, i.e. $\text{HGNN}\left(X_t, G^{H}_t\right)$, is defined as:
\begin{equation*}
    Z_t = \sigma \left(D^{v^{-1/2}}_t H^{H}_t  W^{H}_t D^{e^{-1}}_t H^{H^T}_t D^{v^{-1/2}}_t X_t \Theta\right)
\end{equation*}
where $X_t$ is the feature matrix at time $t$, $\Theta$ is the learnable weight matrix, $D^v_t$ is the node degree matrix at time $t$, $D^e_t$ is the edge degree matrix at time $t$, $W^H_t$ is the diagonal matrix of hyperedge weights at time $t$, and $H^H_t$ is the incidence matrix at time $t$.  

\paragraph{Linear Layer}
The linear layer takes as input $X_t \in \mathbb{R}^{m \times N_s}$, $\widetilde{X}^{TM}_t \in \mathbb{R}^{m \times N_s}$, and $\widetilde{X}^{HG}_t \in \mathbb{R}^{m \times N_s}$ in order to predict the future $q$ values of the target time series $S^{*} \in \mathcal{S}$. \\
First, we concatenate the three inputs:
\begin{equation*}
    \widetilde{X}^{\text{cat}}_t = \left[ \widetilde{X}^{\text{TM}}_t \| \widetilde{X}^{\text{HG}}_t \| X_t \right]
\end{equation*}
Then, the final prediction is obtained through a multi-layer perceptron (MLP):
\begin{equation*}
    \hat{S}_{t+q} =\text{MLP} \left( \widetilde{X}^{\text{cat}}_t \right)
\end{equation*}
where $\hat{S}_t \in \mathbb{R}^q$ represents the predicted output over a forecast horizon $q$.

\section{Numerical Comparisons} \label{section:experimental evaluation}

\subsection{Dataset and Pre-Processing}
For the purposes of our analysis, we consider three distinct multivariate time series datasets corresponding to different application domains: stock returns prediction, appliances' energy consumption forecasting \citep{appliances_energy_prediction_374}, and air quality prediction \citep{air_quality_360}.\\
For the first dataset, the objective is to predict the future value of the S\&P$500$ index, i.e., the target time series, based on the constituent stocks within the index. We collect daily observations from \newdate{hypergraph_date_start}{13}{07}{2010}\displaydate{hypergraph_date_start} to \newdate{hypergraph_date_end}{07}{07}{2023}\displaydate{hypergraph_date_end}, resulting in $3269$ observations for each of the $500$ index components, downloaded from Yahoo Finance\footnote{\url{https://finance.yahoo.com}}. After removing stocks with missing values and computing logarithmic returns, the dataset consists of $440$ stocks, each with $3268$ observations. To reduce the dimensionality and mitigate computational constraints in terms of memory usage and time complexity, we randomly select $10\%$ of the available stocks 
to construct the final dataset.
The second dataset concerns the prediction of appliances' energy consumption. It comprises temperature and humidity measurements collected every $10$ minutes over a period of approximately $4.5$ months within a residential environment, yielding $19735$ observations. Measurements are obtained from multiple zones within a house, as well as from external environmental conditions, resulting in a total of $28$ features.
The third dataset focuses on air quality prediction in an Italian city. It consists of hourly averaged responses from an array of five metal oxide chemical sensors, collected between March $2004$ and February $2005$, for a total of $9558$ observations. After removing categorical variables and imputing missing values using the most recent available observations, the dataset comprises $15$ features.\\
To effectively capture nonlinear patterns, all input features across the three datasets are normalized prior to model training using a rolling window approach. Specifically, for each feature, the mean and standard deviation are computed over a sliding window of size $w$, and the normalized value is obtained as:
\begin{equation*}
\widetilde{x}^{i}_{t} = \frac{x^{i}_{t} - \mu^{i}_{t}}{\sigma^{i}_{t}},,
\end{equation*}
where $i$ denotes the feature index. This normalization procedure ensures that the data are appropriately scaled while preserving their temporal dynamics.

\subsection{Baseline Models and Evaluation Metrics}
For benchmarking purposes, we consider the following baseline models: a Graph Convolutional Network (GCN) combined with an LSTM layer, an attention-based LSTM (AT-LSTM), an attention-based Gated Recurrent Unit (AT-GRU), an attention-based bidirectional LSTM (AT-BiLSTM), a convolutional neural network augmented with an attention-based LSTM (CNN-AT-LSTM), a Transformer, and the \emph{Time–Geometric} (TG) model \citep{gregnanin2026statistical}. The selection of these baselines is motivated by the need to evaluate the performance gains achieved by incorporating hypergraph structures across different modeling paradigms. In particular, we compare purely temporal models (e.g., AT-LSTM, AT-GRU, Transformer and AT-BiLSTM), hybrid models that combine temporal and pairwise structural information (e.g., CNN-AT-LSTM and GCN with LSTM), and ensemble architectures (e.g., TG). This comparison aims to highlight the importance of capturing higher-order relationships among feature variables, as opposed to relying solely on pairwise dependencies typically modeled through graph-based approaches.\\
For graph-based models, the graph representation is constructed from the correlation matrix computed over the features in each dataset. Specifically, we apply the Planar Maximally Filtered Graph (PMFG) \citep{tumminello2005tool} to filter the correlation matrix, retaining only the most significant correlations while enforcing planarity, i.e., ensuring that the resulting graph can be embedded in a plane without edge crossings. In contrast, for the TG model, the graph representation is derived using the visibility graph algorithm applied to the target time series of each dataset.\\
To evaluate and compare the predictive performance of the considered models, we adopt several standard metrics for time series forecasting, namely Root Mean Squared Error (RMSE), Mean Absolute Error (MAE), and Mean Absolute Percentage Error (MAPE). The formal definitions of these evaluation metrics are provided in Appendix \ref{appx_eval_metrics}.

\subsection{Hyperparameters Setting}
To train the proposed framework, each dataset is partitioned into training, validation, and test sets using a $0.7$-$0.15$-$0.15$ split, respectively. The validation set is employed to mitigate overfitting through the use of an early stopping strategy with a tolerance of $10$ iterations. Model optimization is performed using the Adam optimizer \citep{kingma2014adam}, with the number of training epochs fixed at $100$. The Mean Squared Error is adopted as the loss function, and hyperparameter optimization is conducted using the Optuna Python package \citep{akiba2019optuna}.\\
For the proposed framework, the set of hyperparameters to be optimized includes: the learning rate ($lr$), the $\alpha$ parameter of the ELU activation function ($\alpha$), the number of temporal layers ($\#$ layers), the number of neurons in the temporal layers ($\#tl$), the number of neurons in the hypergraph neural network ($\#h$), the time window size ($w$), the number of attention heads (head), the batch size (BS), and the dropout rate (Dropout). It is important to note that the number of layers in the hypergraph neural network is fixed to two, as increasing the depth leads to a significant degradation in performance \citep{huang2021unignn}. The selected hyperparameter values for each dataset are reported in Table \ref{tab:hyperparameters_hypergraph}.
The hyperparameters of the baseline models are selected following the same optimization strategy, and are reported in Tables \ref{tab:hyperparameters_baseline_gcn}-\ref{tab:hyperparameters_baseline_tg} in Appendix \ref{appx_hyperparameters_hypergraph}.
\begin{table}[t]
\caption[Hypergraph Neural Networks Hyperparameters]{Definition of the hyperparameters.}
\centering
\resizebox{\columnwidth}{!}{
\begin{tabular}{|l|c|c|c|c|c|c|c|c|c|}
\hline
    Dataset & lr & $\alpha$ & \# Layer & \# tl & \#h & w & head & BS & Dropout \\
    \hline
    Stock Returns & $7.545\times 10^{-6}$ & $0.3$ & $1$ & $50$ & $15$ & $200$ & $6$ & $32$ & $0.1$ \\ 
    \hline
    Energy Consumption & $5.815\times 10^{-6}$ & $0.3$ & $5$ & $15$ & $25$ & $130$ & $2$ & $32$ & $0.4$ \\ 
    \hline
    Air Quality & $9.083\times 10^{-5}$ & $0.3$ & $5$ & $35$ & $35$ & $100$ & $4$ & $32$ & $0.2$ \\
    \hline
\end{tabular}}
\label{tab:hyperparameters_hypergraph}
\end{table}

\subsection{Numerical Results}

\begin{table}[t]
\centering
\caption[Comparison of forecasting models for Hypergraph Neural Networks.]{Comparison of forecasting models across datasets and evaluation metrics.}
\label{tab:results_hypergraph}
\renewcommand{\arraystretch}{1.2}
\resizebox{\columnwidth}{!}{\begin{tabular}{|l|c|c|c|c|c|c|c|c|c|}
\hline
\multirow{2}{*}{Model} 
& \multicolumn{3}{c|}{Stock Returns} 
& \multicolumn{3}{c|}{Energy Consumption} 
& \multicolumn{3}{c|}{Air Quality} \\
\cline{2-10}
& RMSE & MAE & MAPE
& RMSE & MAE & MAPE
& RMSE & MAE & MAPE \\
\hline

AT-LSTM          & $0.01310$ & $0.00982$ & $106.52$ & $90.72$ & $53.67$ & $60.47$ & $1.30001$ & $1.00799$ & $96.69$ \\
AT-GRU           & $0.01309$ & $0.00984$ & $104.13$ & $90.78$ & $53.95$ & $61.10$ & $0.96535$ & $0.70244$ & $55.69$ \\
BiAT-LSTM        & $0.01309$ & $0.00983$ & $103.82$ & $90.40$ & $51.14$ & $54.84$ & $1.30002$ & $1.00804$ & $96.70$ \\
CNN-AT-LSTM      & $0.01309$ & $0.00983$ & $103.29$ & $90.79$ & $53.95$ & $61.09$ & $1.16101$ & $0.87183$ & $74.88$ \\
Transformer      & $0.01381$ & $0.01028$ & $255.29$ & $89.98$ & $50.63$ & $54.66$ & $0.69199$ & $0.47588$ & $37.14$ \\
GCN with LSTM    & $\mathbf{0.00991}$ & $\mathbf{0.00726}$ & $\mathbf{103.17}$ & $65.63$ & $31.57$ & $\mathbf{28.82}$ & $0.64949$ & $\mathbf{0.45108}$ & $\mathbf{34.34}$ \\
TG               & $0.01309$ & $0.00983$ & $108.52$ & $90.39$ & $50.88$ & $54.24$ & $1.29985$ & $1.00610$ & $96.18$ \\
HGNN             & $0.01316$ & $0.00981$ & $146.35$ & $\mathbf{61.03}$ & $\mathbf{29.85}$ & $29.41$ & $\mathbf{0.64901}$ & $0.45745$ & $34.98$ \\
\hline
\end{tabular}}
\end{table}
Table \ref{tab:results_hypergraph} reports the numerical results obtained across all the considered datasets. The best value for each evaluation metric within each dataset is highlighted in bold.
From the table, it can be observed that the HGNN model does not consistently outperform the baseline models; rather, its performance depends strongly on the characteristics of the dataset. In highly stochastic settings, such as the stock returns dataset, the proposed framework exhibits relatively weaker performance. A plausible explanation is that, in such contexts, higher-order relationships contribute less to the predictive task, while pairwise dependencies are more informative. This interpretation is supported by the superior performance of the GCN combined with an LSTM, which explicitly models pairwise interactions.
Conversely, for datasets where higher-order relationships are more likely to be present - such as energy consumption and air quality - the proposed HGNN framework achieves competitive results. In the air quality dataset, the model attains the lowest values for two out of three evaluation metrics, while in the energy consumption dataset it achieves the best result for one metric and remains close to the best performance in the others. These findings suggest that hypergraph-based representations are particularly effective when complex, non-pairwise dependencies exist among the variables.
It is also important to highlight the poor performance of the \emph{Time-Geometric} model. This can be attributed to the construction of the graph representation via the visibility graph using a feature selected at random from the available variables. This design choice was made to prevent potential data leakage and avoid introducing bias. In particular, constructing the graph directly from the target variable would provide the model with information that should not be available at prediction time, thereby compromising the validity of the evaluation.

\section{Conclusion} \label{section:conclusion}
In this chapter, we introduced a geometric deep learning framework designed to model and exploit higher-order relationships within time series. We proposed a novel methodology for deriving hypergraph representations from multivariate time series without relying on prior structural knowledge. In conjunction with this representation, we introduced a Hypergraph Neural Network (HGNN) framework to effectively capture and leverage higher-order dependencies among variables.\\
From an empirical perspective, we assessed the performance of the HGNN model across three distinct time series datasets: stock returns, energy consumption, and air quality. The results indicate that the proposed framework achieves strong performance for datasets where higher-order relationships are expected, such as energy consumption and air quality. Conversely, its performance deteriorates in highly stochastic settings, such as stock returns, where higher-order dependencies appear to be less informative than the pairwise relationship.

\bibliographystyle{plain}
\bibliography{references}
\newpage

\appendix

\section{Random Matrix Theory} \label{Appx:rmt}
We consider the Random Matrix Theory (RMT) as principal approach for filtering similarity matrices. By filtering, we refer to the process of removing noisy or spurious elements from a similarity matrix while retaining only the most informative relationships among time series.\\ 
The RMT aim is to extract meaningful information from the similarity matrix by distinguishing statistically significant eigenvalue components from noise. Consider a correlation matrix computed from $N$ independent random time series of length $T$. Under the asymptotic regime $N\rightarrow +\infty$, $T\rightarrow +\infty$, with $1 < \lim T/N < +\infty$, the eigenvalues of the correlation matrix follow the Marcenko-Pastur distribution \citep{laloux1999noise,plerou1999universal}, given by: 
\begin{equation}
    \rho(\lambda) = \begin{cases} \frac{Q}{2 \pi\sigma^2}\frac{\sqrt{(\lambda_{+}-\lambda)(\lambda-\lambda_{-})}}{\lambda} &  \text{if } \ \lambda \in [\lambda_{-},\lambda_{+}]\,, \\
    0 & \text{otherwise}
    \end{cases}\label{eq:rmt_distribution} 
\end{equation}
and zero otherwise. Here, 
\begin{equation*}
    Q=\lim \frac{T}{N}\,, \ \ \quad \ \ \lambda_{\pm}=\sigma^{2}\left(1\pm \sqrt{\frac{1}{Q}}\right)^2\,,
\end{equation*}
and $\sigma^{2}$ denotes the variance of the matrix elements, often estimated empirically as:
\begin{equation*}
    \sigma^{2} = 1 - \frac{\lambda_{max}}{N}
\end{equation*}
where $\lambda_{max}$ is the largest eigenvalue of the correlation matrix. Within the RMT framework, eigenvalues exceeding $\lambda_{+}$ are considered statistically significant, while those within the Marcenko–Pastur bounds are attributed to noise. Accordingly, any correlation matrix can be decomposed into the sum of a structural component, denoted by $C^{(s)}$, which is associated with eigenvalues exceeding the threshold $\lambda_{+}$, and a noise component, denoted by $C^{(r)}$, that can be expressed as:
\begin{equation}
C^{(r)}=\sum_{i:\lambda_i \leq \lambda_{+}}\lambda_iv_iv_i^{\dagger}\,. \label{eq:noise_component}
\end{equation}
where $v_i$ denotes the eigenvector associated with eigenvalue $\lambda_i$, and $v_i^\dagger$ its conjugate transpose. \\
In empirical financial data, an eigenvalue significantly larger than all others is typically observed and is commonly referred to as the market mode \citep{sinha2010econophysics,mehta2004rmt,laloux1999noise}. This dominant component reflects the collective market behavior affecting all assets. Removing the market mode is therefore essential for enhancing the detection of meaningful inter-asset correlations. The correlation matrix can thus be decomposed as:
\begin{equation}
    C = C^{(r)} + C^{(m)} + C^{(g)}\,, \label{eq:correlation_decomposition1}
\end{equation}
where
\begin{align}
    C^{(m)}&=\lambda_{max}v_{max}v_{max}^{\dagger}\,, \label{eq:market_component}\\
    C^{(g)}&=\sum_{i:\lambda_+ < \lambda_i < \lambda_{max}}\lambda_iv_iv_i^{\dagger}\,, \label{eq:real_component}
\end{align}
Here, $C^{(r)}$ represents the noise component, $C^{(m)}$ the market component, and $C^{(g)}$ the remaining structurally significant correlations. In this chapter, the matrix $C^{(g)}$ is used as the filtered correlation matrix.

\section{Community Detection} \label{appx_modularity}
Consider a network composed of $N$ nodes, represented by an adjacency matrix $A\in \mathbb{R}^{N\times N}$.
The network structure can be analyzed through community detection, which aims to identify groups of nodes that are more densely connected to each other than to the rest of the network. A widely used measure to quantify the structure of the community is modularity. Given a partition of the node set into communities, modularity $Q(\eta)$ is defined as:
\begin{equation}
    Q(\eta) = \frac{1}{2|E|}\sum_{u,v \in V} \left( A(u,v) - \frac{k_u k_v}{2|E|}\right)\delta (\eta_u,\eta_v) \quad \in [-0.5,1]
    \label{eq:modularity}
\end{equation}
where $k_u$ and $k_v$ denote the degrees of nodes $u$ and $v$, respectively, $\eta_u$ denotes the community assignment of node $u$, and $\delta (\cdot,\cdot)$ is the Kronecker delta function, which is equal to $1$ when $\eta_u = \eta_v$ and $0$ otherwise. Modularity measures the extent to which the observed density of edges within communities exceeds that expected under a null model that preserves the degree distribution. High modularity values indicate a strong and well-defined community structure.\\
Community detection aims to identify groups of nodes within a network that are more densely interconnected with each other than with nodes belonging to other groups \citep{barabasi2016network,fortunato2010community}. To identify non-overlapping communities, we employ a modularity optimization framework \citep{newman2004finding}. This approach is selected due to its foundation on a null model that provides a reference structure against which the observed network organization can be evaluated. Modularity serves as a quantitative measure of the quality of a network partition. Specifically, partitions characterized by high modularity exhibit dense intra-community connections and sparse inter-community connections. \\
Consider a network composed of $N$ nodes, represented by an adjacency matrix $A\in \mathbb{R}^{N\times N}$. The objective is to identify non-overlapping communities described by a $N$-dimensional vector $\eta$, where the component $\eta_i$ specifies the community membership of node $u$, as described in \citep{MacMahon2015}. Recall that equation \eqref{eq:modularity} defines the modularity function. More generally, the modularity can be expressed with respect to an arbitrary null model as:
\begin{equation}
    Q(\eta) = \frac{1}{A_{tot}}\sum_{i,j}\left[A_{ij} - \langle A_{ij} \rangle \right]\delta(\eta_i,\eta_j)\,, \label{eq:modularity_null_model}
\end{equation}
where $\delta(\eta_i, \eta_j)$ is the Kronecker delta function, equal to $1$ if $\eta_i=\eta_j$ and $0$ otherwise, ensuring that only node pairs belonging to the same community contribute to the sum. The normalization constant $A_{tot}=\sum_{i,j}A_{ij}=2l$ corresponds to twice the total number of links $l$ in the network, and $\langle A_{ij} \rangle$ denotes the expected adjacency under the selected null model. A commonly adopted null model is the configuration model\footnote{Used to define the modularity function in equation \eqref{eq:modularity}.}, where $\langle A_{ij} \rangle = \frac{k_ik_j}{2l}$, with $k_i$ representing the degree of node $i$ \citep{fortunato2010community}.\\
When financial networks are constructed from finite time series data, the corresponding correlation matrix often exhibits a dominant global component, commonly referred to as the market mode. In this setting, modularity can be reformulated by incorporating the RMT decomposition of the correlation matrix. Specifically, the modularity can be expressed as:
\begin{align}
    Q(\eta) &= \frac{1}{C_{norm}}\sum_{i,j}\left[C_{ij} - C^{(r)}_{ij}-C^{(m)}_{ij} \right]\delta(\eta_i,\eta_j) \,, \nonumber \\ 
    & = \frac{1}{C_{norm}}\sum_{i,j} C^{(g)}_{ij}\delta(\eta_i,\eta_j)\,. \label{eq:modularity_RMT}
\end{align} 
where $C^{(r)}$, $C^{(m)}$, and $C^{(g)}$ denote, respectively, the noise, market, and significant correlation components\footnote{Defined in equations \eqref{eq:noise_component}, \eqref{eq:market_component}, and \eqref{eq:real_component}.}. The normalization constant is given by $C_{norm}=\sum_{i,j}C_{ij}$. Previous studies \citep{utsugi2004rmt,potters2005rmt,plerou2002rmt} have shown that the eigenvectors associated with $C^{(g)}$ often exhibit alternating sign structures, enabling the identification of groups of assets influenced by common latent factors. This property provides a robust basis for community detection in financial networks.

\section{Numerical Comparisons}

\subsection{Evaluation Metrics} \label{appx_eval_metrics}
The following standard evaluation metrics: Mean Square Error (MSE), Root Mean Square Error (RMSE), Mean Absolute Error (MAE), and Mean Absolute Percentage Error (MAPE) are defined as follows:
\begin{align}
    \text{MSE} &= \frac{1}{l}\sum_{i=1}^{l}(y_i-\hat{y}_{i})^{2}\,,   \label{eq:MSE_evalmetric} \\
    \text{RMSE} &= \sqrt{\frac{1}{l}\sum_{i=1}^{l}(y_i-\hat{y}_{i})^{2}}\,,  \label{eq:RMSE_evalmetric} \\
    \text{MAE} &= \frac{1}{l}\sum_{i=1}^{l}|y_i-\hat{y}_{i}|\,,  \label{eq:MAE_evalmetric} \\ 
    \text{MAPE} &= \frac{1}{l}\sum_{i=1}^{l}\left|\frac{y_i-\hat{y}_{i}}{y_i}\right|\,, \label{eq:MAPE_evalmetric}
\end{align}
where $l$ denotes the number of samples in the corresponding dataset split, $y_i$ is the true target value, and $\hat{y}_i$ is the model prediction.

\subsection{Baseline Models Hyperparameters}
\label{appx_hyperparameters_hypergraph}

For the empirical analysis, we consider the following baseline models: a Graph Convolutional Network (GCN) augmented with a Long Short-Term Memory (LSTM) layer, an attention-based LSTM model (AT-LSTM), an attention-based Gated Recurrent Unit (AT-GRU), an attention-based bidirectional LSTM (AT-BiLSTM), a convolutional neural network augmented with an attention-based LSTM (CNN-AT-LSTM), a Transformer, and the \emph{Time–Geometric} (TG) model. \\
The common hyperparameters across all deep learning models include the batch size (BS), the time window size ($w$), the learning rate ($lr$), and the dropout rate (Dropout). For the GCN–LSTM model, the specific hyperparameters are: the number of GCN layers ($\#$ Layer), the number of neurons in the GCN component ($\#$ tg), and the number of neurons in the LSTM layer ($\#$ tl). The corresponding optimized hyperparameter values are reported in Table \ref{tab:hyperparameters_baseline_gcn}.

\begin{table}[h]
\caption[GCN–LSTM Baseline Model Hyperparameters]{Definition of the hyperparameters for the GCN-LSTM Model.}
\centering
\resizebox{\columnwidth}{!}{
\begin{tabular}{|l|c|c|c|c|c|c|c|}
\hline
    Dataset & BS & w & lr & Dropout & $\#$ Layer & $\#$ tg & $\#$ tl  \\
    \hline
    Stock Prices & $7.545\times 10^{-6}$ & $0.3$ & $1$ & $50$ & $15$ & $200$ & $6$  \\ 
    \hline
    Energy Consumption & $5.815\times 10^{-6}$ & $0.3$ & $5$ & $15$ & $25$ & $130$ & $2$  \\ 
    \hline
    Air Quality & $9.083\times 10^{-5}$ & $0.3$ & $5$ & $35$ & $35$ & $100$ & $4$  \\
    \hline
\end{tabular}}
\label{tab:hyperparameters_baseline_gcn}
\end{table}

For the attention-based LSTM model, the specific hyperparameters include: the number of LSTM layers ($\#$ Layer), the number of neurons in the LSTM ($\#$ tl), and the number of attention heads. The optimized configurations for this model are provided in Table \ref{tab:hyperparameters_baseline_atlstm}. Note that also AT-GRU, AT-BiLSTM, and CNN-AT-LSTM models have the same specific hyperparameters set and for simplicity we consider the same values.

\begin{table}
\caption[AT–LSTM, AT-GRU, AT-BiLSTM, and CNN-AT-LSTM Baseline Models Hyperparameters]{Definition of the hyperparameters for the AT-LSTM, AT-GRU, AT-BiLSTM, and CNN-AT-LSTM Models.}
\centering
\resizebox{\columnwidth}{!}{
\begin{tabular}{|l|c|c|c|c|c|c|c|c|c|}
\hline
    Dataset & BS & w & lr & Dropout & $\#$ Layer & $\#$ tl \\
    \hline
    Stock Prices & $7.545\times 10^{-6}$ & $0.3$ & $1$ & $50$ & $15$ & $200$  \\ 
    \hline
    Energy Consumption & $5.815\times 10^{-6}$ & $0.3$ & $5$ & $15$ & $25$ & $130$  \\ 
    \hline
    Air Quality & $9.083\times 10^{-5}$ & $0.3$ & $5$ & $35$ & $35$ & $100$ \\
    \hline
\end{tabular}}
\label{tab:hyperparameters_baseline_atlstm}
\end{table}

For the \emph{Time–Geometric} model, the hyperparameters depend on the configuration of both the temporal and geometric components. In this study, we adopt an LSTM architecture for the temporal component and a GCN for the geometric component. Accordingly, the specific hyperparameters include: the number of layers in the LSTM ($\#$ Layer LSTM), the number of neurons in the LSTM ($\#$ tl), the number of layers in the GCN ($\#$ Layer GCN), the number of neurons in the GCN ($\#$ tg), and the number of neurons in the fully connected layers ($\#$ tc). The optimized hyperparameter values for the TG model are reported in Table \ref{tab:hyperparameters_baseline_tg}.

\begin{table}
\caption[\emph{Time–Geometric} Baseline Model Hyperparameters]{Definition of the hyperparameters for the \emph{Time-Geometric} Model.}
\centering
\resizebox{\columnwidth}{!}{
\begin{tabular}{|l|c|c|c|c|c|c|c|c|c|}
\hline
    Dataset & BS & w & lr & Dropout & $\#$ Layer LSTM & $\#$ tl & $\#$ Layer GCN & $\#$ tg & $\#$ tc \\
    \hline
    Stock Prices & $7.545\times 10^{-6}$ & $0.3$ & $1$ & $50$ & $15$ & $200$ & $6$ & $32$ & $0.1$ \\ 
    \hline
    Energy Consumption & $5.815\times 10^{-6}$ & $0.3$ & $5$ & $15$ & $25$ & $130$ & $2$ & $32$ & $0.4$ \\ 
    \hline
    Air Quality & $9.083\times 10^{-5}$ & $0.3$ & $5$ & $35$ & $35$ & $100$ & $4$ & $32$ & $0.2$ \\
    \hline
\end{tabular}}
\label{tab:hyperparameters_baseline_tg}
\end{table}


\end{document}